# Synthesis of electrical networks interconnecting PZT actuators to damp mechanical vibrations.

Francesco dell'Isola[1], Edmund G. Henneke[2], and Maurizio Porfiri[3]

[1] *Dip. Ingegneria Strutturale e Geotecnica, Università di Roma "la Sapienza", 00184 Roma, Italia*

[2,3] *Dept. Engineering Sciences and Mechanics, "Virginia Tech" 24060 Blacksburg VA USA*

**Abstract**. This paper proves that it is possible to damp mechanical vibrations of some beam frames by means of piezoelectric actuators interconnected via passive networks. We create a kind of electromechanical wave guide where the electrical velocity group equals the mechanical one thus enabling an electromechanical energy transfer. Numerical simulations are presented which prove the technical feasibility of proposed device.

1.  **Introduction**

In recent years there has been an increasing interest in the structural control by electric devices, exploiting the piezoelectric effect. The standard approach to the problem (see for instance [1]) makes use of lumped devices, requiring high actuators performances and inductors which are usually not realizable by means of passive electric circuits, because of their large inductance. Furthermore quoted approach enables the designer to control only one vibration mode of the structure, since the lumped circuit is resonant only at a chosen frequency mode of the structure. In [2], [3], [4] an axially and transversely vibrating Euler beam is coupled to a second order transmission line connecting a series of piezoelectric actuators distributed along the beam; the so obtained synergy of actuators succeeds in lowering the optimal coupling inductance of the circuit modules, which may make easier the practical realization of inductors. Moreover, the *Elastica* and the telegraph equation show different dispersive relation, then it is possible to obtain an optimal coupling only for a fixed wave-length thus limiting the range of applications of the device to a one-mode control.

In this paper we show that it is possible to damp mechanical vibrations in a given structure, constituted by Euler beams, by means of a set of piezoelectric actuators mechanically connected to the beams and interconnected via electrical technically feasible *passive* networks. Since the most efficient way to damp mechanical vibrations by means of electrical networks, is to achieve a self-resonant modal coupling between the electrical and the mechanical motions (that is to design an electric circuit with the same dispersion relation as the *Elastica*) a distributed circuit analog to the Euler beam has to be synthesized. Indeed we will find a passive circuit the equation of which is governed by a discrete version of the *Elastica* equations and on which we can impose the boundary conditions usually imposed in frames.

---

[1] Francesco dell'Isola: francesco.dellisola@uniroma1.it, *corresponding author*.
[2] Edmund G Henneke: henneke@vt.edu
[3] Maurizio Porfiri: mporfiri@vt.edu

We will solve this synthesis problem applying the black box approach to mechanical systems, developed by many engineers and scientists during the 1940's in an attempt to design analog computers [5], [6]. Indeed, in section 2, considering a transversely vibrating Euler beam, we will subdivide the entire domain of the beam into a finite number of equal small beam elements. Then regarding each beam element as a two-terminal black box we will provide its mobility matrix representation relating, in Laplace transforms, the velocities at the terminals to the contact actions. Hence, supposing the size of the beam element to be sufficiently small, we will consider for this mobility matrix a Laurent expansion with respect to the length truncated the at a suitable degree of approximation. Finally we will synthesize a four-port passive circuit module simulating the beam element at the terminals and cascade connecting the modules we will find an electric circuit analog to the entire beam. Once the electric analog is designed, in section 3 we will properly take into account the piezoelectric effect in the circuit modules and in the *Elastica* equation finding the governing equations of the so called piezo-electromechanical beam. In section 4 we illustrate the previous results by means of numerical simulations; a purely flexible aluminum beam simply supported and electrically insulated is chosen as *ad hoc* example time-evolution problem. The performances of the piezo-electromechanical beam will be discussed, proving the technological feasibility of the conceived circuit issued by the low performances requested to the circuit elements constituting a circuit module, and the efficiency of the conceived device. Indeed, it allows for a quick and complete energy exchange between its mechanical and electrical forms, using available piezoelectric actuators.

## 2. Electromechanical analogies

### 2.1 The black box approach

It is well known that there is a wide class of physical devices whose interaction with the outer world can be fruitfully modeled by the exchange of energy only at a finite number of access points, called *terminals*; the state of every terminal being exhaustively described by a finite number of state variables.

This kind of approach is usually referred to as the *black box* approach, and, as we will see, it allows for a complete discussion of a large class of electro-mechanical devices leading to an extension of concepts deeply rooted in the Theory of Networks to the Dynamics of Structures. Furthermore, the black box approach provides theoretical tools and design criteria to synthesize electrical networks analog to mechanical structures. We assume that the state of each terminal $T_i$ is fully characterized by a pair of *l*-tuples

$$(a_i, \tau_i) = \left((a_i^l, \ldots, a_i^l), (\tau_i^l, \ldots, \tau_i^l)\right) \tag{1}$$

We will refer to them, respectively, as to *across* and *through* variables and we will assume the inner product of a through and an across variable to have the dimensions of a power. Considering electrical networks the across variables are voltage drops between the terminals and a reference voltage, while the currents flowing into the network at the terminal are regarded as through variables. When dealing with an *n*-port network **N** [7] we will group the port voltages into *n*-column vector $\boldsymbol{v}$ and the port currents into *n*-column vector $\boldsymbol{i}$. For purely mechanical structures the state at each terminal is described by a 3-column vector of velocities regarded as across variable and a 3-column vector of contact actions regarded as through variable. In fact, considering a planar structure and an orthonormal basis $(\boldsymbol{e_1}, \boldsymbol{e_2}, \boldsymbol{e_3})$ the across variable is constituted by the velocity $v_1$ along $e_1$, the velocity $v_2$ along $e_2$ and the angular velocity $\omega$ about $e_3$, while the through variable is formed by the $t_1$ component of the contact force, the $t_2$ component of the contact force and the $\boldsymbol{e_3}$ component of the contact bending moment M. Given an *n*-terminal planar structure $\Sigma$ we will group the velocities into a 3*n*-column vector $w$ and the contact actions into a 3*n*-column vector $\boldsymbol{\alpha}$. A Piezoelectric actuator cannot either be modeled as a network or as mechanical structure, since the state of each terminal needs to be characterized by both mechanical and electrical descriptors. We will refer to such a device as an electromechanical structure and in section 3, when dealing with the piezo-electromechanical beam, we will provide its mathematical model.

Generally the so called conjugate variables $(a_i, \tau_i)$ are supposed to belong to the *Signal Space* $D_+^l$ [8]; this choice is driven by the physical need to consider a space general enough to model the signals to be considered and is embedded in Schwartz distribution space [8] [9], so that powerful and useful representation theorems are available. In order to mathematically model the different behaviors of distinct physical devices, for instance networks or mechanical devices, we have to specify for a fixed device its set of allowable conjugate variables. For example a resistor compels the potential difference between its terminals to be proportional to the current flowing in it, an inductor compels the voltage drop between its terminals to be proportional to the time-derivative of the current flowing in it and a spring compels the difference of velocities between its terminals to be proportional to the time derivative of the applied force.

Then given a κ-terminals physical device $\mathcal{D}$ the binary relation $\mathcal{C}_S$ on $D_+^{l \times k}$ is said to characterize $\mathcal{D}$ if the set of admissible signals is:

$$S = \{(\boldsymbol{\alpha}, \boldsymbol{\tau}) \epsilon D_+^{l \times k} \times D_+^{l \times k}, \boldsymbol{\alpha} \mathcal{C}_S \boldsymbol{\tau}\} \tag{2}$$

Considering a physical device $\mathcal{D}$ completely solvable, linear and time-invariant [4] it is possible to fully characterize $\mathcal{D}$ by means of a univocally determined *impulsive response*; furthermore, supposing all the signals to be Laplace transformable [6] we can introduce, whenever it is possible, the *immittance* and *transmission* matrix representations of the device in the frequency domain. In particular we will call impedance of an *n*-port network **N**, the $n \times n$ matrix **Z** defined by:

$$V(s) = Z(s)I(s) \tag{3}$$

where the capital letters refer to Laplace transformed variables and $s$ is the Laplace variable. Analogously we will call mobility matrix of an n-terminal structure **Σ**, the $3n \times 3n$ matrix **M** defined by:

$$W(s) = M(s)A(s) \tag{4}$$

Given an *n*-terminal structure **Σ** represented by a mobility matrix **M** we will say that a 3*n*-port network **N** is analog to **Σ**, if the dimensionless mobility matrix of the structure is equal to the dimensionless impedance matrix of the circuit.

## 2.2 Synthesis of the electric analog to the Euler beam

Consider the dimensionless constitutive and balance equations for a purely flexible Euler beam:

$$\frac{\partial \zeta}{\partial \varepsilon} = \vartheta, \quad F_M = \frac{\partial \vartheta}{\partial \varepsilon} \tag{5}$$

$$\frac{\partial F_T}{\partial \varepsilon} - \frac{\lambda l^4 \partial^2 \zeta}{t_0^2 k_M \partial \tau^2} = 0, \quad \frac{\partial F_M}{\partial \varepsilon} + F_T = 0 \tag{6}$$

with $F_M$, $F_T$ dimensionless shear contact action and bending moment; $\zeta, \vartheta$ dimensionless deflection and change of attitude; $\varepsilon, \tau$ dimensionless material abscissa and time; $\lambda, l, k_M, t_0$ indicate, respectively, the density per unit length of the beam, the beam length, the bending stiffness and a typical time which will be fixed according to the particular application we are interested in. Subdividing the entire domain of the beam into *n* equal beam elements of dimensionless size $\delta$, it is easy to achieve the $4 \times 4$ mobility matrix representation of a generic beam element relating the dimensionless contact actions at the two terminals to the dimensionless velocities. The generic entry of the mobility matrix has the following form:

$$M_{ij}(\delta) = \frac{h_{i,j}(\varepsilon)}{-1+\cosh(\sqrt{\overline{\beta}}k\delta)\cos(\sqrt{\overline{\beta}}k\delta)} \tag{7}$$

where η is a holomorhic function, κ is related to the dimensionless Laplace variable $\eta$ by:

$$k = \sqrt{\eta}e^{j\frac{\pi}{4}} \tag{8}$$

and

$$\beta^2 = \frac{\lambda l^4}{t_0^2 k_M} \tag{9}$$

Thus the mobility matrix is Laurent expandable in the neighborhood of the size equal to zero and the ring of convergence of the series is determined by the denominator of (7) which limits the *bandwidth-size* product as follows:

$$\delta\sqrt{\varpi} \leq \frac{4.73004}{\sqrt{\beta}} \tag{10}$$

where $\varpi$ is the dimensionless angular frequency. This condition provides an upper boundary for the size of the mesh defined on the beam once we choose the frequency range of the analog circuit, i.e. the highest mode of vibration we want to control. Truncating the Laurent expansion at a suitable degree of approximation we can express the mobility matrix in the Foster's canonic form [7] as:

$$\boldsymbol{M} = \frac{1}{\eta}\boldsymbol{K_0} + \eta \boldsymbol{K_\infty} \tag{11}$$

which immediately leads to the design of a circuit analog to the beam element.

In fact the mobility matrix can be synthesized by series connecting two networks, the former simulating the pole at zero and the other the pole at infinity, i.e. the capacitive effect and the inductive effect. Each of the networks can be synthesized, following [7], as uncoupled impedances, determined by the eigenvalues of the residue matrix, terminating an ideal transformer, the turns-ratio matrix of which is determined by the eigenvectors of the residue matrix. The values of the turns-ratio matrices linearized in $\delta$ are:

$$T_0 = \frac{\kappa_0}{\sqrt{2}}\begin{pmatrix} 1 & 0 & 1 & 0 \\ -\frac{\delta}{2} & 1 & \frac{\delta}{2} & 1 \end{pmatrix} \qquad T_\infty = \frac{\kappa_\infty}{\sqrt{2}}\begin{pmatrix} \delta/6 & -1 & \delta/6 & 1 \\ 1 & \delta/6 & 1 & -\delta/6 \\ -3\delta/34 & 1 & 3\delta/34 & 1 \\ 1 & 3\delta/34 & -1 & 3\delta/34 \end{pmatrix} \tag{12}$$

With $\kappa_0$ and $\kappa_\infty$ arbitrary real constants. While the impedances terminating the transformers, once we choose for instance $C_1$, are determined by the following set of relations:

$$\frac{C_1}{C_2} = \frac{12}{\delta^2}, \quad C_1 L_2 = \frac{1}{720}\varsigma^2, \quad C_1 L_4 = \frac{1}{4080}\varsigma^2 \quad C_2 L_1 = \frac{1}{48}\varsigma^2, \quad C_2 L_3 = \frac{17}{1680}\varsigma^2 \tag{13}$$

with

$$\varsigma^2 = \left(\frac{\kappa_0}{\kappa_\infty}\right)^2 \frac{\lambda}{k_M}(\delta l)^4 \tag{14}$$

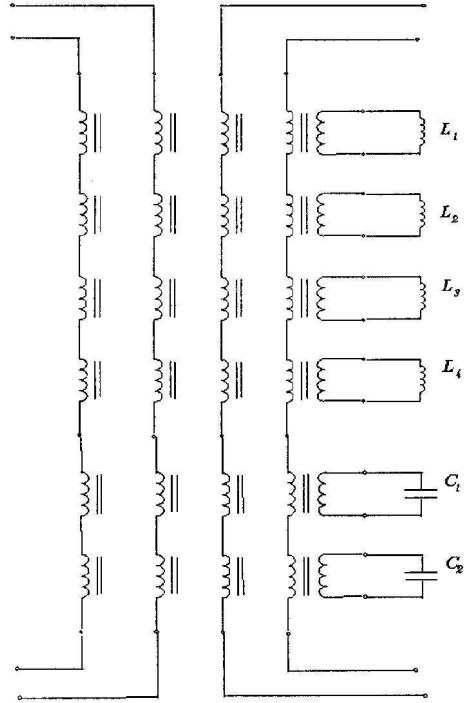

Fig. 1 Circuit-module

## 2. Piezo-electromechanical beam

A bending piezoelectric actuator is a two terminal electromechanical structure, governed by the following constitutive equation:

$$\begin{pmatrix} M \\ Q \end{pmatrix} = \begin{pmatrix} K_{mm} & K_{me} \\ K_{em} & K_{ee} \end{pmatrix} \begin{pmatrix} \chi \\ V_p \end{pmatrix} \qquad (15)$$

where $\chi$ represents the relative change of attitude at the two terminals, $Q$ the capacitive charge stored in the actuator.

Furthermore, assuming the actuator to be lossless it is easy to prove that:

$$K_{mm} \geq 0, \quad K_{ee} \geq 0, \quad K_{em} = -K_{me} \qquad (16)$$

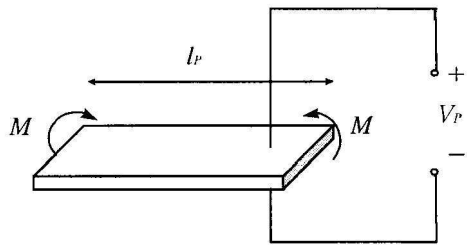

Fig. 2 Piezoelectric bending actuator

The constitutive equation, in particular, establishes that from an electric point of view the actuator behaves like a capacitor in parallel connection with a current source driven by the mechanical term $\chi$

Consider now an Euler beam, of length *l*, and suppose *n* continuously distributed bending actuators to be glued on it; if *n* is sufficiently big, we can imagine these actuators to form a thin piezoelectric layer on the beam.

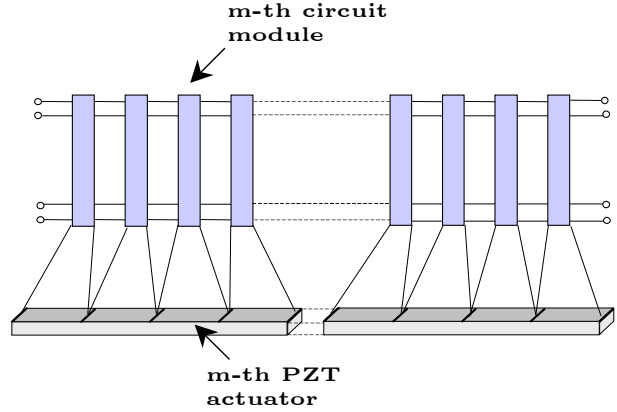

Fig. 3 A scheme of the piezo-electro-mechanical beam

Furthermore, consider the electric circuit analog to the transversally vibrating Euler beam and suppose the *m-th* actuator to be connected as the capacitor $C_1$ of the *m-th* module of the circuit.

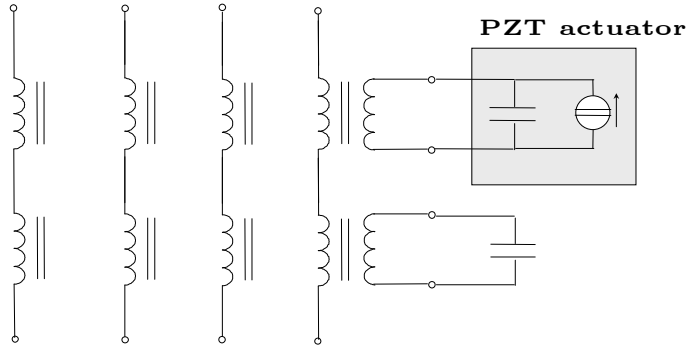

Fig. 4 Actuator connection in the circuit module

The piezoelectric direct effect modifies the evolution equation for the electric distributed circuit in:

$$\varphi''''(\varepsilon,\tau) - \rho\dot{v}''(\varepsilon,\tau) = -\beta^2\ddot{\varphi}(\varepsilon,\tau) \qquad (17)$$

with $\varphi$ dimensionless voltage, $v$ dimensionless velocity and choosing the characteristic voltage to be related to the radius of gyration (characteristic length for the deflection) by the following:

$$V_0 = \frac{\kappa_0}{\sqrt{2}}\frac{r_0}{t_0}\sqrt{\frac{\lambda}{k_{ee}}} \qquad (18)$$

with

$$k_{ee} = \frac{K_{ee}}{l_p} \qquad (19)$$

representing the capacitance per unit length and

$$\rho = \frac{K_{em}l^2}{t_0 k_M}\sqrt{\frac{\lambda}{k_{ee}}} \tag{20}$$

standing for the coupling parameter.

From a mechanical point of view we assume the piezoelectric indirect effect to modify the constitutive equation of the bending moment in:

$$M(x,t) = (k_{mm} + k_M^m)u''(x,t) + K_{me}V_p(x,t) \tag{21}$$

where

$$k_{mm} = K_{mm}l_p \tag{22}$$

is the bending stiffness of the piezo-layer on the PZT beam while $k_M^m$ is the original bending stiffness of the beam. Usually for available actuators the stiffness of the piezo-layer can be neglected compared to the proper bending stiffness of the beam. The voltage drop across the actuator is proportional to the voltage across the first port of the circuit module, the one governed by the elastica equation in the coarse model, in dimensionless variables the relationship can be expressed by:

$$\varphi_p = -\frac{\sqrt{2}}{\kappa_0}\varphi \tag{23}$$

Then the evolution equation for the beam becomes:

$$\upsilon''''(\varepsilon,\tau) + \rho\dot{\varphi}''(\varepsilon,\tau) = -\beta^2\ddot{\upsilon}(\varepsilon,\tau) \tag{24}$$

We explicitly remark that as the capacitance per unit length decreases the coupling effect increases, nevertheless it is worthwhile to use actuators endowed with high capacitance to keep low the inductances and the turning-ratio matrix entries of the interconnecting circuit, so as to dispense with active circuits.

### 4. Modal analysis

Consider a simply-supported piezo-electromechanical beam, where the electric boundary conditions, equivalent to the mechanical ones, are obtained short-circuiting the upper ports and open-circuiting the lower ports of the edge-modules of the circuit (see figures 1, 3). Considering the evolution equation of the system in terms of the vertical displacement and of the time-integral of the voltage, let us expand the solution on the basis functions constituted by the eigenfunctions of the free vibrating simply supported beam. The evolution equations for the Fourier's coefficients for the *m-th* mode will be independent on the evolution of the other modes, avoiding undesired spill-over between different modes. The following picture exhibits the modulation provided by the piezoelectric effect on the free vibrations of the system, supposing that initially all the energy is stored in its mechanical potential form, i.e. the beam is initially at rest and all the electric devices are discharged.

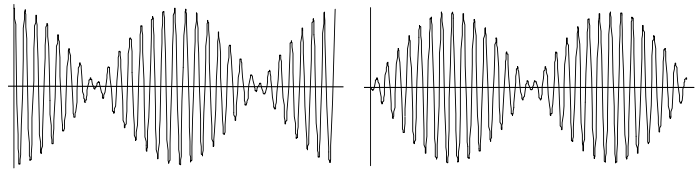

Fig. 5 Fourier's coefficients of a generic mechanical and electrical modes

## 5. Conclusions

This paper has demonstrated the efficiency of the distributed approach in the problem of controlling mechanical vibrations exhibiting an electromechanical device able to control every mode of the structure and suitable to industrial applications, issued by the passivity of all the circuit elements constituting the analog circuit. Moreover, for an aluminum beam 1 meter in length, 3 centimeter in width and 2 millimeter in thickness the analog circuit, with ten modules and available PZT actuators, utilizes inductances less than 10mH, capacitances less than 100nF and the entries of the turning-ratio matrices ranges between 0.1 and 10. Furthermore, the modulation of the Fourier's coefficients shown in figure 5 provide a complete energy transfer from the mechanical into the electrical form in a time equal to eight times the modal vibration period of the beam. The amplitude of the potential difference between the terminals of the actuator for initial deflection in the first mode less than one per cent of the beam length remain always under the maximum voltage level provided by the PZT producer.

**Acknowledgements**

F. d. I. and M. P. wish to thank the Department of Engineering Science and Mechanics of the Virgina Polytechnic Institute and State University for the warm hospitality and research grants provided in the last two years.